# Decoding European Palaeolithic art: Extremely ancient knowledge of precession of the equinoxes

**Martin B. Sweatman\*[1] and Alistair Coombs[2]**

[1]School of Engineering, University of Edinburgh, Mayfield Road, Edinburgh, UK. EH9 3JL
[2]Department of Religious Studies, University of Kent, Canterbury, Kent, UK. CT2 7NF

**Keywords:** Precession, zodiac, Lascaux, Chauvet, Altamira, Lion-man, Göbekli Tepe, Çatalhöyük

## 1. Summary

A consistent interpretation is provided for Neolithic Göbekli Tepe and Çatalhöyük as well as European Palaeolithic cave art. It appears they all display the same method for recording dates based on precession of the equinoxes, with animal symbols representing an ancient zodiac. The same constellations are used today in the West, although some of the zodiacal symbols are different. In particular, the Shaft Scene at Lascaux is found to have a similar meaning to the Vulture Stone at Göbekli Tepe. Both can be viewed as memorials of catastrophic encounters with the Taurid meteor stream, consistent with Clube and Napier's theory of coherent catastrophism. The date of the likely comet strike recorded at Lascaux is 15,150 ± 200 BC, corresponding closely to the onset of a climate event recorded in a Greenland ice core. A survey of radiocarbon dates from Chauvet and other Palaeolithic caves is consistent with this zodiacal interpretation, with a very high level of statistical significance. Finally, the Lion Man of Hohlenstein-Stadel, circa 38,000 BC, is also consistent with this interpretation, indicating this knowledge is extremely ancient and was widespread.

## 2. Introduction

This work concerns our understanding of the astronomical knowledge of ancient people. This knowledge, it seems, enabled them to record dates, using animal symbols to represent star constellations, in terms of precession of the equinoxes. Conventionally, Hipparchus of Ancient Greece is credited with discovering this phenomenon. We show here that this level of astronomical sophistication was actually known at least 36 thousand years earlier.

Evidence accumulated from many ancient archaeological sites, representing dates from at least 38,000 BC to the middle of the Neolithic, overwhelmingly supports this view. The statistical exercise described in Section 6 proves, in a scientific sense, that this view is correct. The list of sites covers many of the most well-known archaeological finds in Europe and the Near East, including;

- The Lion Man of Hohlenstein-Stadel, southern Germany circa 38,000 BC
- Chauvet, northern Spain circa 34,000 BC
- Lascaux, southern France circa 15,000 BC
- Altamira, northern Spain circa 14,000 BC
- Göbekli Tepe, southern Turkey circa 10,000 BC
- Çatalhöyük, southern Turkey circa 7000 BC

The key to cracking this ancient code is provided by the Vulture Stone at Göbekli Tepe, constructed at the Palaeolithic-Neolithic boundary in southern Anatolia. In previous work (1) it was shown how this ancient megalithic pillar can be viewed as a memorial to the proposed Younger Dryas event (2), a collision with cometary debris also recorded by a platinum 'spike' in a Greenland ice core (3) at 10,940 BC, which likely triggered the Younger Dryas period, with all its catastrophic consequences.

\*Author for correspondence (martin.sweatman@ed.ac.uk).

The next clue to this ancient code is provided by Neolithic Çatalhöyük, also in southern Anatolia. We show how animal symbolism at this ancient site can be interpreted using the same method and zodiac as at Göbekli Tepe. It appears we continue to use the same zodiacal constellations today in the West, although some of them are no longer represented by animal symbols and a few of the remaining animal symbols have switched places.

The same method and zodiac can also be used to decode much of the animal symbolism displayed by European Palaeolithic cave art, from the Aurignacian Lion Man of Hohlenstein-Stadel, southern Germany, through to Magdelanean Altamira in northern Spain. The final piece of the logic puzzle is provided by the famous Shaft Scene at Lascaux, which has an almost identical interpretation to the Vulture Stone at Göbekli Tepe. They differ only in the date of the catastrophe memorialized and the recorded radiant of the cometary strike.

This exercise in decoding is entirely logical and quite simple. Like a crossword puzzle, solving one problem provides a clue to the next. Therefore, we begin our account with a brief summary of published findings at Göbekli Tepe and especially Pillar 43, the Vulture Stone.

## 3. Decoding Göbekli Tepe

In previous work (1, 4) several stone pillars at Göbekli Tepe, an ancient hill-top site probably constructed after the Younger Dryas event and before the so-called Neolithic revolution, circa 10,000 BC, were decoded. Pillar 43 provided the statistical key for this interpretation; it is our 'Rosetta Stone' (see Figure 1). Essentially, Pillar 43 can be viewed as a memorial to the proposed Younger Dryas event (5). The date carved into the Vulture Stone is interpreted to be 10,950 BC, to within 250 years, in very good agreement with the platinum spike in the Greenland ice core (3).

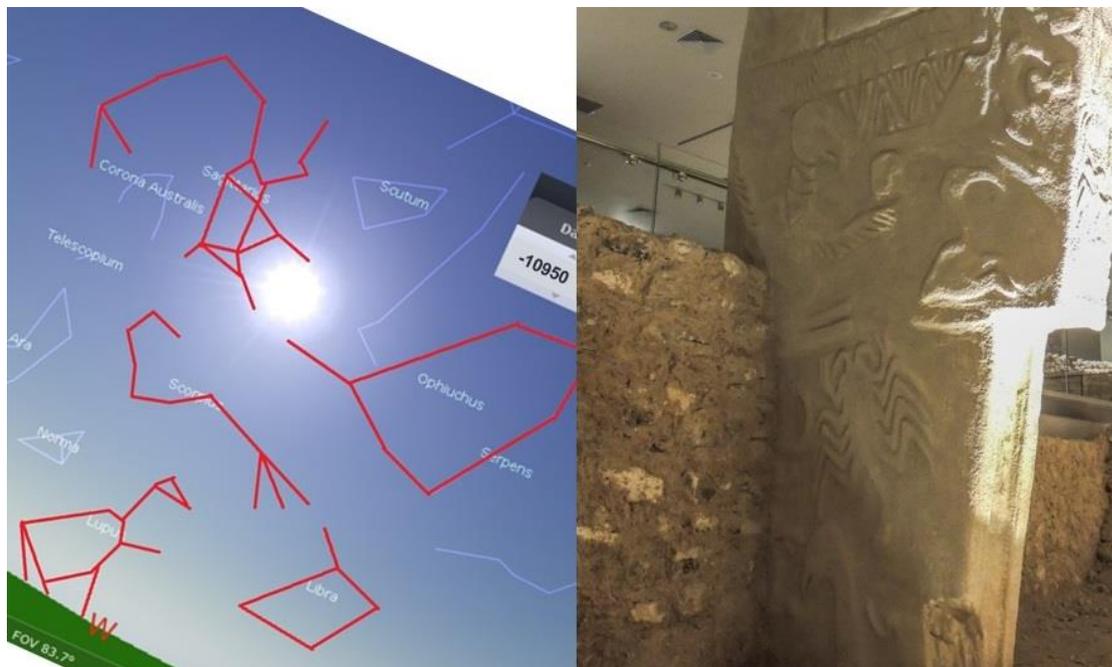

**Figure 1.** Comparison of Pillar 43 (copy in Sanliurfa museum) with constellations around Scorpius (left image from Stellarium).

Furthermore, Pillar 2 at Göbekli Tepe can be interpreted as the path of the radiant of the Taurid meteor stream at the time the site was occupied, and Pillar 18 can be interpreted as indicating the Younger Dryas event was caused by an encounter with Taurid meteor stream debris from the direction of northern Aquarius, in accordance with Clube and Napier's theory of coherent catastrophism (6-8). The animal symbols decoded from these three pillars are listed in Table 1. The probability that these pattern matches are purely coincidental is estimated to be around 1 in 300,000 (see Appendix B), which is not quite low enough to claim a scientific discovery. Again, this estimate relies on matching animal symbols to their corresponding constellations, and is therefore open to criticism of subjectivity.

The statistical case supporting this view is based on an analysis of the probability that the animal symbols on Pillar 43 could have appeared in their respective positions by pure chance, given they match their associated star constellations so well. Given there are several pillars at Göbekli Tepe with repeated animal symbols, a statistical estimate in the region of 1 in 100 million that the animal patterns on Pillar 43 could have occurred by pure chance is obtained (see Appendix A). This estimate is based on ranking the animal symbols against each potential constellation, shown in Table 1, and is therefore open to criticisms of subjectivity. To dispute this statistical case, one would need to argue that the ranking shown in Table 1 is significantly flawed, and that for each associated constellation there are several animal symbols at Göbekli Tepe that provide a better fit than the ones that actually appear on Pillar 43.

## 4. Decoding Çatalhöyük

Çatalhöyük is thought to be the first Neolithic town in Southern Anatolia, with maximum population perhaps as much as 8000 (9). Radiocarbon dating has established that its lowest occupation layers date to around 7250 BC. It appears to have been largely destroyed by an intense fire around 6400 BC, with later occupation layers dating to around 6250 BC on the eastern site. The younger western site was likely occupied by a different culture, considering that symbolism on pottery and methods of construction are quite different, between 6200 and 6000 BC. Çatalhöyük is therefore several millennia younger than Göbekli Tepe, forming a bridge in time between the date represented by Pillar 43 and the Bronze Age.

Many different types of animal motif appear at Çatalhöyük, from boar tusks to bear claws, expressed either as paintings, wall 'inclusions', or 'installations'. The most prominent and significant, by far, are the many installations found in rooms interpreted to be religious shrines by the site's archaeologists. These consist of large wall and floor features that appear to have been re-plastered and re-painted every year. Only four types of these large installation are known, each appearing frequently in Çatalhöyük shrine rooms, corresponding to the following animals; aurochs, ram, leopard and another symbol that has been interpreted as either a goddess figure or a bear (9). So far, the reason why only these specific animals are represented in shrines, and therefore the basis of their religion, is unknown. We show here that their symbolism is identical to that displayed at Göbekli Tepe.

To show that the same symbolic code is used at Çatalhöyük as at Göbekli Tepe we need to locate the corresponding solstices and equinoxes. Taking the representative date 7000 BC corresponding to earlier occupation levels, we find using Stellarium, an archaeoastronomical software, that the constellations corresponding to the solstices and equinoxes are (see Figure 2);
- Summer solstice = Virgo
- Autumn equinox = Capricornus
- Winter solstice = Aries
- Spring equinox = Cancer

If we convert these constellations to the symbols used at Göbekli Tepe, using Table 1, we find;
- Summer solstice = down-crawling quadruped
- Autumn equinox = aurochs
- Winter solstice = unknown
- Spring equinox = unknown

Let's consider these associations in turn. The down-crawling quadruped appears at the top-right of Pillar 43 at Göbekli Tepe (see Figure 3), although the symbol is difficult to identify precisely. Given that we are decoding an ancient form of proto-writing, which would not have used symbols that are too similar to each other to avoid confusion, it is likely that this symbol is the same as a similar symbol on display at Sanliurfa museum, also shown in Figure, recovered from Göbekli Tepe. Now compare with a drawing of a Çatalhöyük shrine room, shown in Figure 4. It is clear a similar symbol appears in this room above the central bucrania, although it is the other-way-up. The face drawn on this symbol at Çatalhöyük is the artist's interpretation – no face can be discerned on the actual installations as they were normally deliberately destroyed when a house was abandoned. However, the circular symbol on the animal's belly is correctly drawn.

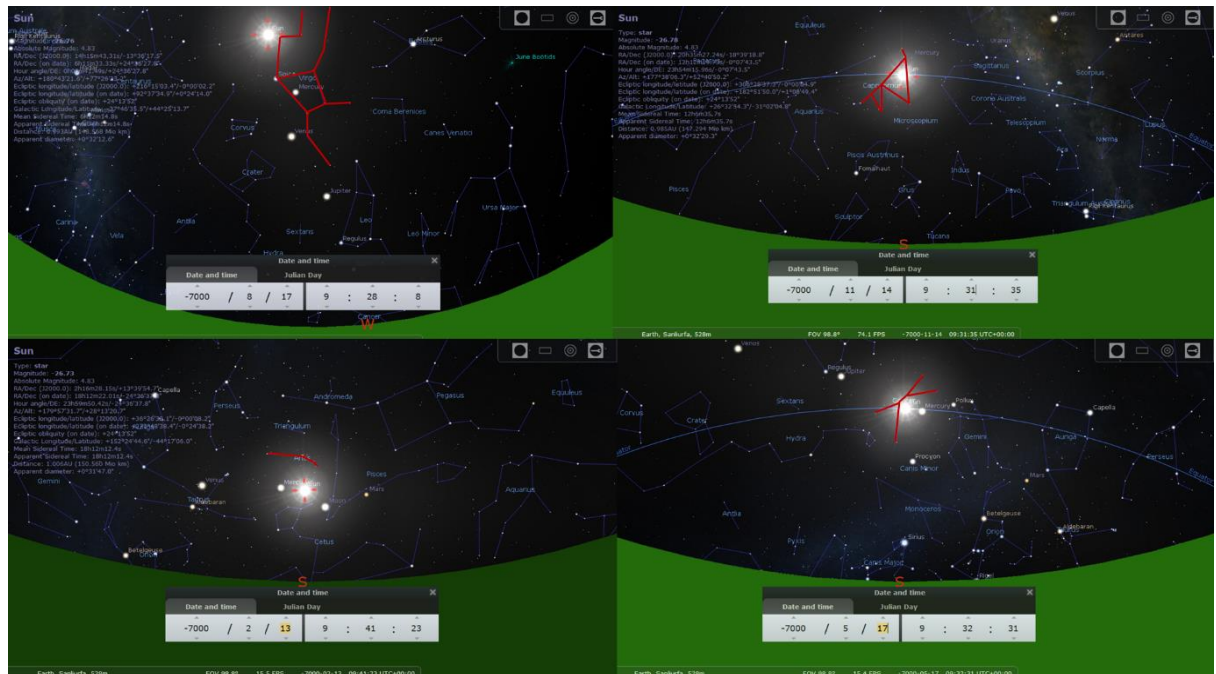

**Figure 2.** Summer solstice (top left), winter solstice (bottom left), spring equinox (top right), and autumn equinox (bottom right), at 7000 BC, southern Anatolia in 7000 BC, corresponding to Virgo, Aries, Capricornus and Cancer respectively (images from Stellarium).

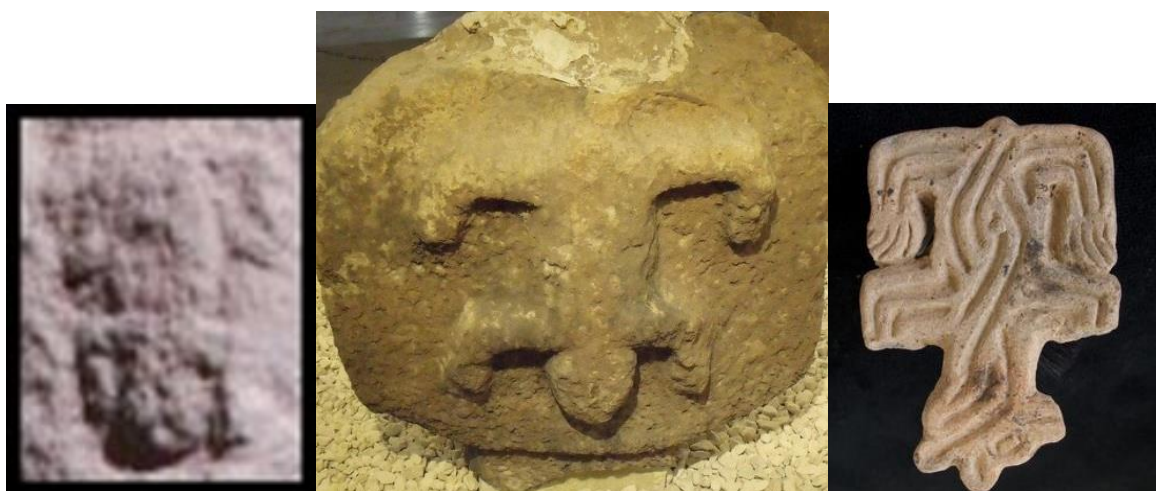

Figure 3. Comparison of ancient Anatolian bear symbols. Left: the symbol on the top-right of Pillar 43 at Göbekli Tepe. Middle: a symbol on display at Sanliurfa museum, recovered from Göbekli Tepe. Right: bear seal stamp found at Çatalhöyük (from www.Çatalhöyük.com).

This cultic symbol has caused some confusion among the site's excavators. The site's original excavating director in the 1960s, James Mellaart, described it as a Goddess symbol, with splayed legs, perhaps pregnant and giving birth (10). This contributed to the development of a Goddess Cult focussed on the site (11). With more recent excavations, directed by Ian Hodder, it has become clear this symbol is probably a splayed bear with a stubby tail (9). This is because a seal stamp, or similar item, has been discovered at Çatalhöyük with the same overall profile, but also with sufficient details to identify it as a bear – see Figure 3.

According to our interpretation, this symbol should represent the summer solstice, and therefore the circle on its belly likely represents the mid-day sun, just like the circle on Pillar 43 above the vulture/eagle's wing at Göbekli Tepe. Therefore, the 'down-crawling quadruped' identified at the top-right of Pillar 43 is now clearly identified as a bear. Table 1 is updated to reflect this.

The aurochs symbol at Göbekli Tepe appears at the top of Pillar 2, and has been interpreted to indicate the constellation Capricornus (1). Therefore, we should find aurochs installations in Çatalhöyük shrine rooms, this time representing the autumn equinox. And indeed, we see several bucrania in the Shrine Room, shown in Figure 4. Indeed, bucrania are some of the most common installations in Çatalhöyük shrine rooms, indicating a special reverence for this particular constellation, possibly because of its earlier association with the Taurid meteor stream.

According to our interpretation, we should also find installations representing Aries in Çatalhöyük shrine rooms. Unfortunately, we have yet to identify the animal symbol representing Aries. Several animal symbols found at Göbekli Tepe have yet to be associated with any constellation, and are therefore candidates. Given today's association of Aries with the Ram, which appears at Göbekli Tepe on Pillar 1, Enclosure A, as well as on at least one other pillar at Göbekli Tepe, it is tempting to make this association. And indeed, a ram installation is apparent in the Çatalhöyük Shrine Room (see Figure 4). This strongly suggests that the Ram = Aries also at Göbekli Tepe. Like Aries, other constellations with their associated animal symbols also appear to have survived the millennia to modern times, such as the scorpion (Scorpius) and dog/wolf (Lupus, see Table 1).

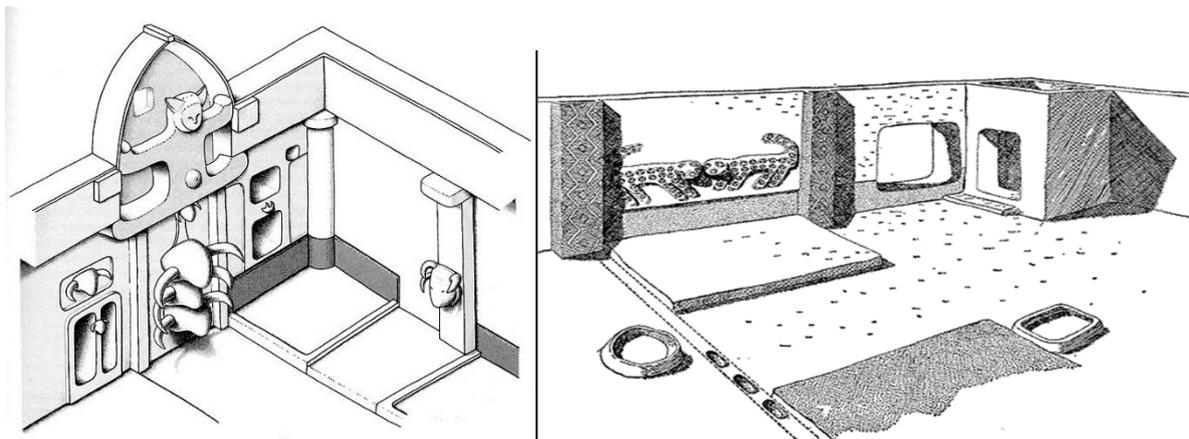

Figure 4. Artists impression of Shrine Rooms at Çatalhöyük (from (10)). Left: a shrine room with aurochs bucraniums, rams heads, and a bear symbol. Right: a shrine room with twin leopards.

Finally, according to our interpretation, we should seek symbolism associated with Cancer. Today's symbol, the Crab, is unknown at either Göbekli Tepe or Çatalhöyük. Therefore, this is likely a more modern association, and we should instead seek another animal symbol prominent at Çatalhöyük, that also appears at Göbekli Tepe, but has yet to be associated with any constellation and would provide a good fit to the Cancer constellation. The only remaining installation type at Çatalhöyük is the leopard. Leopard symbolism at Çatalhöyük appears in several prominent locations, including an installation with a pair of leopards facing each-other in another shrine room (see Figure 4). At Göbekli Tepe, a lion or leopard appears on Pillar 51, Enclosure H, and has yet to be linked to any constellation. Moreover, Cancer at sunset can be viewed as a leopard or lion pouncing or running. Indeed, at Çatalhöyük twin leopards are found facing each other, further emphasizing the symmetry of the Cancer constellation. We therefore suggest it is likely that leopard or lion symbolism represents Cancer. It is tempting to narrow this association to leopards only, but this is not yet known with certainty. It might well have been the case that Cancer was represented by any large feline Therefore, animal symbolism at Çatalhöyük is perfectly consistent with our interpretation of Göbekli Tepe, and we have been able to deduce two new animal symbols: Aries = ram and Cancer = large feline. These new animal symbols are listed in Table 2.

## 5. Decoding the Lascaux Shaft Scene

The caves at Lascaux are famous across the world for their remarkable Palaeolithic cave art. In reality, they are just one particularly splendid example among many different caves in Europe. Indeed, Chauvet is even more extraordinary given its extreme age, being around 20,000 years older than Lascaux and yet displaying a similar level of artistry.

Dating of the Lascaux cave system is uncertain. Estimates range from around 17,000 to 13,000 BC. Unfortunately, it has not been possible to radiocarbon date the art itself because its pigments are not organic. The animals displayed at Lascaux are very similar to those displayed at the Neolithic sites discussed above. But there also a few additions, including many horses, several stags, and a single rhinoceros in the Shaft Scene.

The key to decoding Lascaux, and therefore other Palaeolithic art, is interpretation of the Shaft Scene. This well-known scene is quite separate from all the other artwork at Lascaux, being situated at the bottom of a deep shaft, suggesting it has a special status. It is also unique among Palaeolithic artworks in that it depicts a man, apparently falling in a manner suggesting injury or death – see Figure 5.

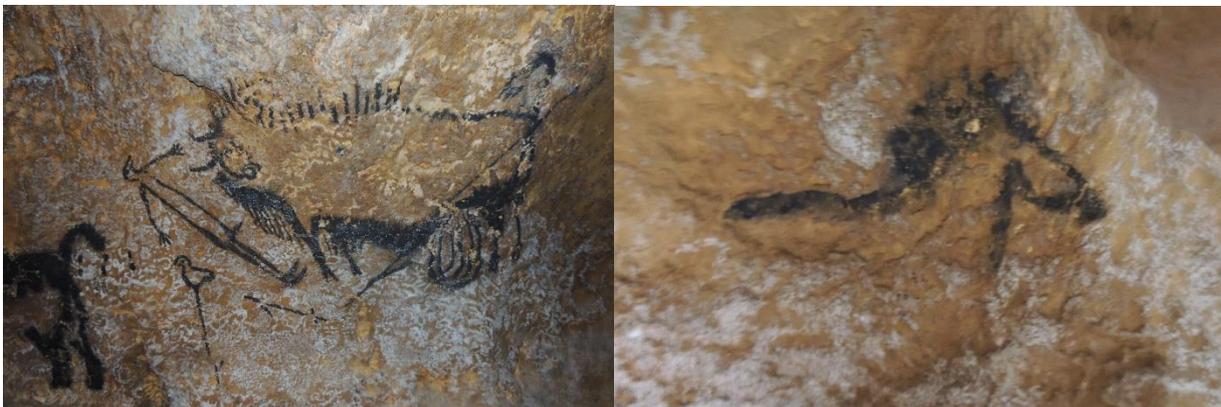

Figure 5. The Lascaux Shaft Scene. Left: main panel with rhino, duck/goose and disembowelled aurochs/bison with dying man on the main wall. Right: horse on rear wall.

Another clue to the meaning of the Shaft Scene is provided by the fact that only four different animal symbols are displayed here; a bison/aurochs, duck/goose, and rhinoceros (to the left of the falling/dying man) on the main wall with a horse on the rear wall. The bison is particularly striking, apparently pierced by a spear. It also seems to be dying, given its entrails are hanging underneath. The horse on the back wall is not often described as being part of this scene, but it is central to the interpretation described next.

Similarities with Göbekli Tepe's Vulture Stone are striking. Both display a man, possibly dead or dying and both display four prominent animal symbols. At Göbekli Tepe the four animals are the vulture/eagle, bear, ibex/gazelle and tall bending bird corresponding to the four solstices and equinoxes at the date of the Younger Dryas event. It is therefore sensible to enquire whether the Shaft Scene at Lascaux is equivalent to the Vulture Stone of Göbekli Tepe and can therefore be decoded using the same method.

Noting the bison/aurochs and duck/goose symbols in the Shaft Scene, and using Table 1 and Stellarium we immediately find the following;
- Bison/aurochs = Capricornus = summer solstice between 15,350 and 13,000 BC
- Duck/goose = Libra = spring equinox between 15,700 and 14,100 BC

Therefore, this scene might represent a date anywhere between 15,350 and 14,100 BC. To narrow down this range we need to consider the other two animal symbols. Unfortunately, neither of these symbols has previously been decoded. But logically, they are unlikely to correspond to constellations that have already been decoded. When we consider this date range we see the following possibilities;

- Autumn equinox: Taurus 15,350 to 14950 BC, or Aries 14950 to 14,100 BC
- Winter solstice: Leo 15,350 to 14,800 BC, or Cancer 14,800 to 14,100 BC

Given that in Tables 1 and 2, Aries is represented by the ram and Cancer is represented by a large feline, and that rams and felines are recorded in Palaeolithic art, it is likely the date range is limited to between 15,350 and 14,950 BC, and therefore the rhinoceros and horse likely represent Taurus and Leo. When we consider these constellations at sunset (see Table 3), which is the convention for this system (1), we find that the rhinoceros and horse are good fits to their respective constellations (Taurus and Leo), which provides further confidence in this interpretation. We therefore suggest the Shaft Scene encodes the date 15,150 ± 200 BC, and we have now completed our ancient zodiac.

Tables 1, 2 and 3 together list the entire zodiac so far deciphered. As there are a few other animal symbols apparent in Palaeolithic art, such as the deer/megaloceros and the mammoth, it is likely there are some regional and temporal variations of this zodiac that remain to be decoded, but they are not investigated further here.

Now that we have a date, we can try to interpret the scene. What should we make of the falling/dying man and the speared/dying bison? Given that the Vulture Stone at Göbekli Tepe very likely refers to the Younger Dryas event and, according to Napier and Clube's theory of coherent catastrophism, this is unlikely to be an isolated incident, could the Shaft Scene represent another encounter with the Taurid meteor stream?
At Göbekli Tepe, the fox features on the largest central pillars of the largest enclosure yet uncovered, indicating the event dated by the Vulture Stone refers to a cosmic event from the direction of northern Aquarius. Instead, the Shaft Scene displays an injured aurochs, representing Capricornus, not a fox. Is the aurochs here equivalent to the fox at Göbekli Tepe? To answer this, we need to consider the precession of the Taurid meteor stream.

As described earlier (1), the longitude of the ascending node of the Taurid meteor stream is expected to precess at the rate of one zodiacal sign every six thousand years (12). Today, the Taurid meteor stream radiant is centred (and hence maximal) over Aries. Therefore, at the time of the Younger Dryas event, around 13 thousand years ago, it would have been centred over Aquarius, described at Göbekli Tepe in terms of the fox. On the date depicted by the Shaft Scene, around 17 thousand years ago, its centre would have been over Capricornus. Therefore, the injured aurochs in the Shaft Scene is consistent with its interpretation as a Taurid meteor strike from the direction of Capricornus. Hence the injured or dying man might indicate a catastrophic encounter with the Taurids, as for the Vulture Stone of Göbekli Tepe.

Clearly, we should seek independent evidence of a catastrophic comet strike at this time. The Younger Dryas event is known as a millennial-scale climatic fluctuation. Therefore, we should first investigate if there is any strong climatic fluctuation at the time indicated by the Shaft Scene. Figure 6 shows that, very interestingly, there is a fairly strong climatic fluctuation at precisely this time recorded by a Greenland ice core (13). Indeed, when we take into account the fact that the Greenland ice core chronology is expected to differ from the radiocarbon chronology by at least 70 years at this time (14), we can see that the onset of the climatic fluctuation at 15,300 BC agrees very well with our interpretation of the Lascaux Shaft Scene.

Of course, there remains the possibility that this interpretation of the Shaft Scene is wrong, and any similarity with symbols at Göbekli Tepe and Çatalhöyük is coincidence. To this end, in the next section our ancient zodiac and the methodology described here are compared with the known dates of reliable radiocarbon evidence obtained directly from animal symbols and figurines at Palaeolithic cave sites across Europe.

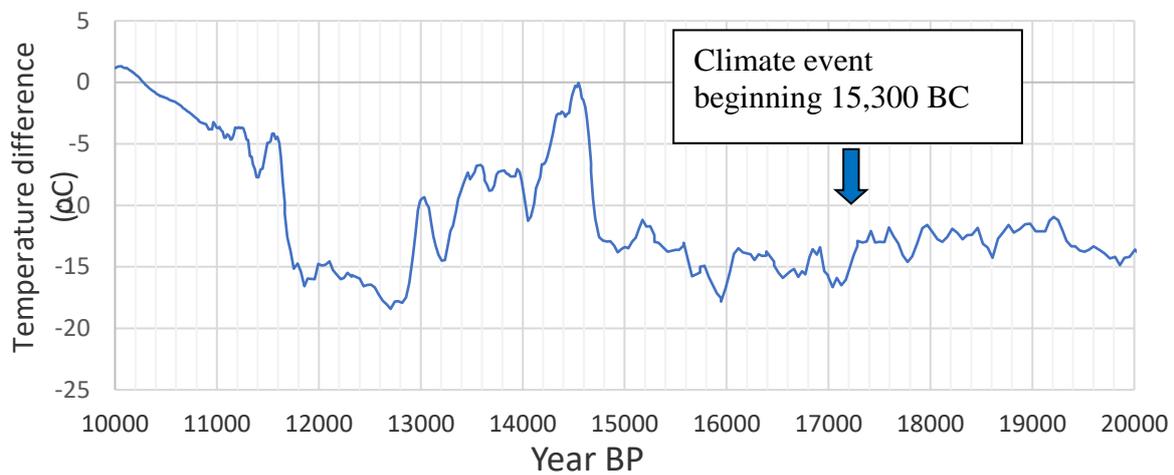

Figure 6. Greenland temperature variation reconstructed from the GISP2 ice core (13). Year BP indicates the number of years before 1950 AD.

## 6. Decoding European Palaeolithic cave art

Thousands of examples of Palaeolithic art have been uncovered across Europe. Unfortunately, relatively few have been dated directly by high quality radiocarbon assays. Nevertheless, there are now a sufficient number of reliable dating exercises published to statistically test the hypothesis of this work. As we are comparing accepted descriptions of these Palaeolithic animal symbols, such as bison, horse etc., with their corresponding published radiocarbon dates, this statistical test is entirely objective and scientific. It does not suffer from any degree of subjectivity.

In the following test, we use *all* reliable radiocarbon dates for Palaeolithic cave art animal symbols published in English-language peer-reviewed research journals (15-22). This includes 24 dates corresponding to animal symbols found in 9 caves in France and Spain, and 2 dates corresponding to zoomorphic figurines found in 2 German caves. We also include the 2 shrines at Çatalhöyük predicted on the basis of the animal symbols decoded from Göbekli Tepe. Table 4 details each entry.

For each animal symbol in Table 4, we find the appropriate solstice or equinox corresponding to that animal, whichever is nearest to the calibrated radiocarbon date. We then determine the 'separation' between the radiocarbon date of the art and the solstice or equinox date of the centre of the corresponding constellation.

Only the most reliable data can be used for this exercise. Therefore, we do not include any samples for which the uncertainty (at 1σ) in the calibrated radiocarbon date exceeds 1074 years (which is 1/3rd of the maximum separation of 3221.5 years between our predictions and the radiocarbon dates). This is because radiocarbon data that exhibits large experimental uncertainty cannot be used to distinguish between a successful prediction and pure chance. Furthermore, for some animal paintings several radiocarbon assays have been performed. Considering that each animal symbol is expected to have been painted by a single artist in one go, we treat these cases as follows. Where multiple radiocarbon assays of the same animal painting agree to within 2 standard errors (2σ) we take their average. Where they do not, there is very likely a problem with one or more of the measurements and so we reject them. Finally, we do not include any data from Cosquer Cave, a coastal cave which is partly below sea level, for paintings that are below the high-tide mark as these paintings are likely contaminated and their radiocarbon dates are therefore unreliable.

If the interpretation of this work is false, there should be no correlation between these specific solstices and equinoxes and the radiocarbon dates of animal figures in Palaeolithic caves, i.e. the measured separation should be evenly distributed across 3,221.5 years (one eighth of Earth's precessional cycle). But, Table 4 and Figure 7 show, in fact, there is an extremely strong correlation. In other words, the radiocarbon dates of the animal symbols listed in Table 4 are highly correlated with the dates of their associated equinoxes and solstices. Considering that each zodiacal constellation, on average, represents ± 25,772 / 24 years, if our hypothesis is correct we should expect a roughly uniform distribution of separations up to 1074 years, tailing off beyond this due to experimental uncertainty in the radiocarbon dates. This is precisely what is observed.

It is possible to estimate the probability that this correlation is due to chance. We see that all samples, except one from Cosquer Cave, have separations of at most 1450 years. The Cosquer outlier has a separation of 1720 years. Considering the expected uniform distribution for the null hypothesis, the probability of this skewed distribution occurring for these 28 samples is 28 x (1720/3221.5) x (1450/3221.5)$^{27}$, which is equivalent to a chance of 1 in 150 million. In a scientific sense, our hypothesis is proven.

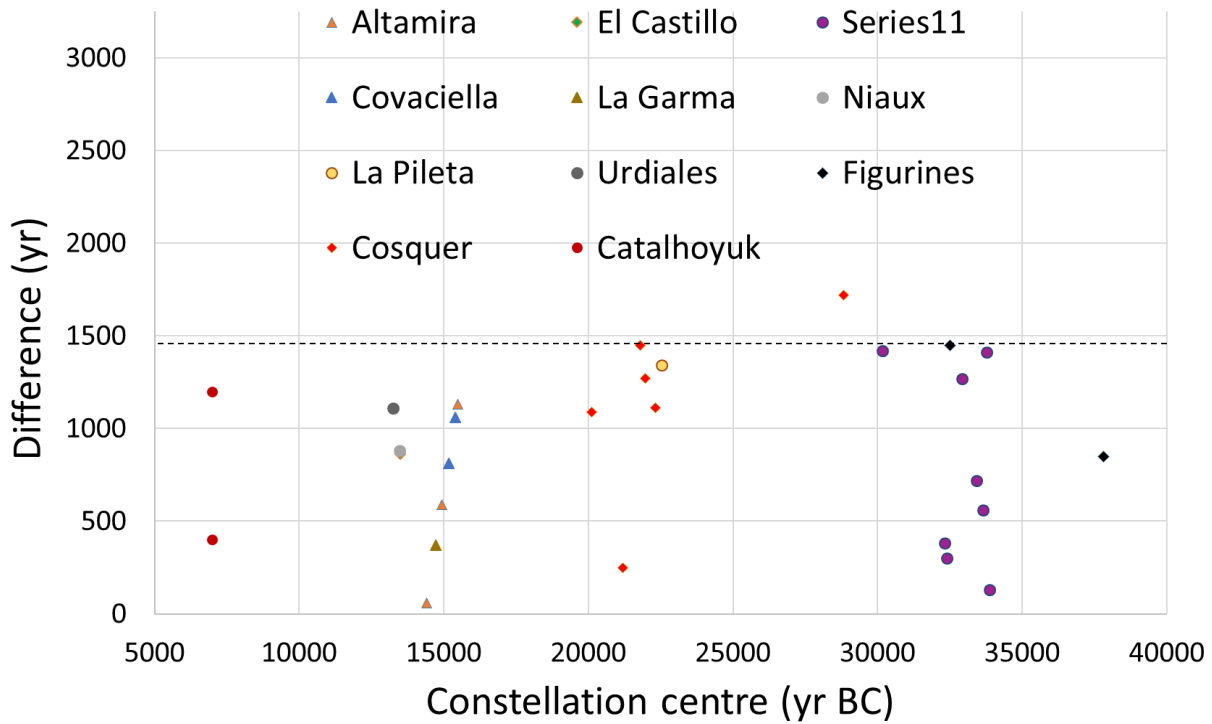

Figure 7. Correlation between the dates of solstice/equinox constellations and the radiocarbon dates of the corresponding animal symbols. Data in Table 4.

## 7. Conclusions

We have defined a zodiac that is consistent with the Lascaux Shaft Scene, Çatalhöyük shrines and Göbekli Tepe using precession of the equinoxes.

When we use it to work out the date of the Lascaux Shaft Scene, we find it is 15,150 BC to within 200 years, which agrees with proposed dates for the paintings at Lascaux. In addition, the wounded bull at Lascaux describes the position of maximum intensity of the Taurids when Lascaux was occupied.

When we use it to work out a date range for when Çatalhöyük was occupied, we find it is 7,400 – 6,500 BC, which agrees with the main occupation phase of Çatalhöyük.

And, when we use it to work out the date of the Vulture Stone at Göbekli Tepe, we find it is 10,950 BC to within 250 years, which agrees with the known date of the Younger Dryas event. Moreover, we also find that Pillar 2 at Göbekli Tepe describes the path of the radiant of the Taurid meteor stream when Göbekli Tepe was occupied, and Pillar 18 describes the position of the maximum intensity of the Taurids.

We get all this from a single zodiac, using precession of the equinoxes. The evidence to support this view, is;

- The probability that the Vulture Stone could match the relevant parts of the sky is extremely tiny, around 1 in 100 million by pure chance according to the rankings in Table 1.
- The probability that the Vulture Stone describes the date of the Younger Dryas event, at the same time that Pillar 2 describes the path of the radiant of the Taurid meteor stream, at the same time that Pillar 18 describes the position of its maximum intensity, is around 1 in 300,000 by pure chance, according to the rankings in Table 1.
- The probability that this zodiac could match by pure chance the radiocarbon dates of Palaeolithic cave art and the 2 shrines at Çatalhöyük is 1 in 150 million.

We emphasize that our final statistical test for the Palaeolithic cave art is completely objective. We have used all the available data that meet our unbiased quality criteria. By itself, the resulting probability of 1 in 150 million of obtaining this data set by pure chance is scientific proof of our hypothesis, which is that animal symbols were used in very ancient times to write dates using precession of the equinoxes.

When we combine the outcome of this statistical test with our statistical estimate of the pattern matches on Göbekli Tepe's Vulture Stone, we find that the probability that all these correlations could have occurred by pure chance is around 1 in 15,000 trillion, which is completely negligible. Even though this particular figure is open to a degree of subjectivity according to the rankings in Table 1, this uncertainty is trivial compared to this overwhelming statistical result. Therefore, we have cracked this ancient zodiacal code.

This code was likely used for many tens of thousands of years until comparatively recently. Its origin, evolution, distribution and ending are presently unknown. Clearly, this zodiac is not fixed, either temporally or geographically. There are likely many local variations, including those that must have occurred between the end of Çatalhöyük's occupation and today. For example, the bull has moved from Capricornus to Taurus and the feline symbol has moved from Cancer to Leo.

Two of the ancient sites discussed here, Göbekli Tepe and Lascaux, appear to both represent specific moments in time that involve catastrophic encounters with the Taurid meteor stream. We saw that the probability that Göbekli Tepe is unrelated to the Younger Dryas event is around 1 in 300 thousand. Given that the Lascaux Shaft Scene also appears to implicate the Taurid meteor stream from the direction of Capricornus, we can reduce this by a factor of 12, the probability of choosing Capricornus at random, to yield a probability of 1 in 4.6 million that the Taurids are not involved. On this basis we claim that Napier and Clube's theory of coherent catastrophism is almost certainly correct. Given the climate oscillation beginning around 15,300 BC (GISP2 chronology), there is good motivation to search for corresponding geochemical evidence for this event.

The theory of coherent catastrophism predicts such events should not be isolated. We should, therefore, enquire whether other examples of fine Palaeolithic art, such as Chauvet, signal further such encounters. For example, recent work that analysed megafaunal remains in Alaskan and Yukon 'muck' indicates similar events, of unknown scale, likely occurred around 18, 30, 37, 40 and 48 thousand years ago (23).

Finally, it appears the capabilities of ancient people have been severely underestimated, at least as far as astronomy is concerned. The level of astronomical knowledge uncovered here at such an ancient time calls into question standard models of diffusion and migration of humans in general. For instance, if ancient people could also estimate longitude via the lunar method, a not unreasonable expectation for someone with knowledge of precession of the equinoxes, then they could have navigated the oceans as soon as sufficiently robust vessels could be built. Indeed, the potential 'impact' of the Taurid meteor stream on the evolution, dispersal and development of mankind, and other animals, in general appears to require some revision.

# Appendix A

It is worth revisiting the statistical case for Pillar 43. This pillar has 8 animal symbols that we suggest represent known constellations. Considering that other pillars at Göbekli Tepe display multiple versions of the same animal symbol, the total number of different animal symbol combinations on Pillar 43 is simply $12^8$ = 430 million, since there are 12 animal symbols in Table 1. According to our ranking of the animal symbols in Table 1, the probability of choosing at random, assuming all the different combinations are equally likely, a set of animal symbols as good as the one that actually appears on the pillar is just 2 in 430 million. However, the three small animal symbols at the top of the pillar could be read left-to-right or right-to-left. Therefore, we are now at a probability of 4 in 430 million.

There are two more factors we need to take into account, that roughly cancel. The first involves the presence of the scorpion. Rather than finding the probability that any combination of animal symbols on pillar 43 could match any region of the sky, in terms on the western constellation set in Stellarium, we have so far found the probability that any combination of animal symbols on Pillar 43 could match the region of the sky that surrounds Scorpius. As this is a more restrictive premise, our current probability estimate is too low. To obtain a better estimate we can simply eliminate the Scorpion from consideration. In other words, we take the position of the scorpion as being fixed. We therefore need to multiply by a factor of 12, giving 48 in 430 million.

However, our analysis of permutations of animal symbols on Pillar 43 has so far not considered the probability of their precise positioning, given a specific combination, on the pillar. For example, the angle subtended by the dog/wolf – scorpion – bird triplet is very similar to the angle subtended by Lupus – Scorpius – Libra in the sky. To take account of this strong correlation we can divide Pillar 43 into several regions within which only one animal pattern can appear (see Figure A1). Here, we are mainly interested in that part of the pillar where there is some freedom to choose the position of the animal symbols, i.e. on the main part which constitutes four animal symbols surrounding the scorpion. By dividing this part of the pillar into 8 regions surrounding the scorpion, each region defines an arc of 45 degrees. We suppose that the four animal symbols around the scorpion could have appeared in any of these 8 regions, providing their clockwise order is fixed. As it is, they appear to be in almost exactly the correct orientation around Scorpius to match the positions of the constellations relative to the ecliptic and the setting sun, except that the bending bird with down-wriggling fish (which we match to Ophiuchus) is about 45 degrees (i.e. one region) out of place. It should be in region 3, not 2, in Figure A1. We ask, what is the probability that these 4 symbols could match the correct positions of the constellations by pure chance?

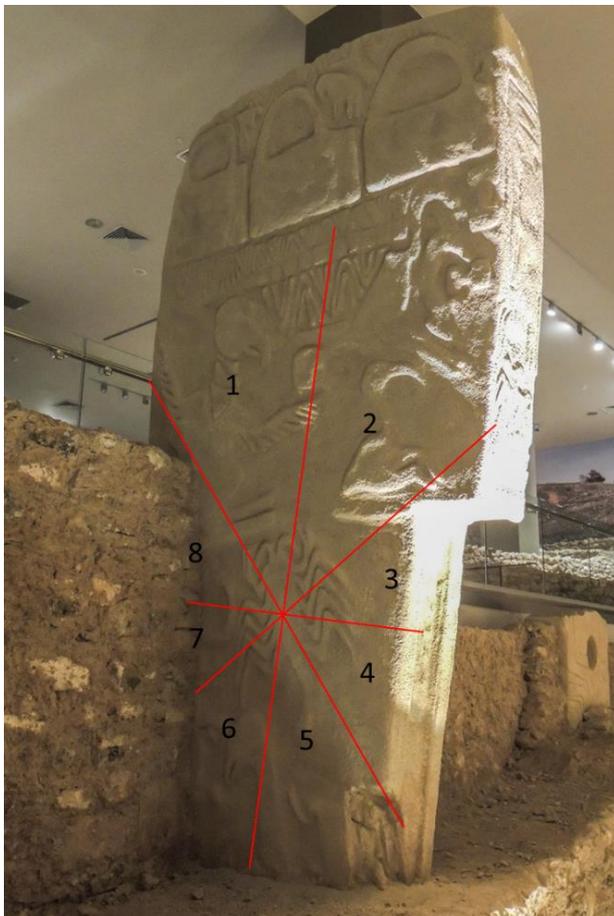

Figure A1. Division of Pillar 43 into regions to enable analysis of the spatial correlation of its main panel.

If we define the first region chosen as region 1, with the remaining regions labelled 2 to 8 clockwise, then the remaining 3 animal symbols can be placed, in clockwise order, in regions 2, 5, 6, or in 3, 5, 6 or in 2, 5, 7 or in 2, 4, 6. Any of these 4 situations could be deemed to be as good, or better, than the one that actually occurs on Pillar 43, as they are all wrong by at most one position. The total number of configurations available without changing the clockwise order of the animal symbols is 5 + (4 x 2) + (3 x 3) + (2 x 4) + 5 = 35. Therefore, the good orientational ordering of these 4 animal symbols on the main part of Pillar 43 around the scorpion has a chance of around 4 in 35 of occurring by pure chance. This nearly cancels the factor of 12 that comes from eliminating the scorpion. So, we are back to 4 in 430 million, or around 1 in 100 million. This is the chance the animal symbols on Pillar 43 could have matched their respective constellations by pure chance. To dispute this, one would need to make the case that there are at least 100, and not just 2, different combinations of animal symbols that are equally as good as the combination that actually appears on the pillar.

# Appendix B

There are three potential coincidences at Göbekli Tepe that we need to address. Multiplying their probabilities together, presuming their independence, will provide an overall estimate of the probability they could have occurred together;

1. The date written on the Vulture Stone is extremely close to the accepted date of the Younger Dryas impact event.
2. Pillar 2 describes the path of the radiant of the Taurid meteor stream, the same meteor stream thought to be responsible for the Younger Dryas impact event according to Clube and Napier's theory of coherent catastrophism.
3. Pillar 18, the central dominant pillar of Enclosure D, refers to the northern portion of Aquarius, which would have been at the centre of the Taurid meteor stream, its point of maximum intensity, at the time.

Let's consider each point in turn.

1. The earliest radiocarbon date for Enclosure D at Göbekli Tepe is for the mortar of the rough stone wall, at 9530 BC to within a few hundred years. The date written on the Vulture Stone is 10,950 BC to within a few hundred years, while the Younger Dryas event, according to a Greenland ice core occurred at 10,940 BC to within 10 years, which is about 10,870 BC according to the radiocarbon chronology. The chance of finding a date on the Vulture Stone that is within 100 years of the Younger Dryas event date, and yet is over 1,400 years before the earliest accepted radiocarbon date, is about 100/1400 = 1 in 14.
2. Pillar 2 has the sequence (crane, fox, aurochs) representing the Taurid radiant path (Pisces, northern Aquarius, Capricornus). There are $12^3$ = 1728 different possible animal symbol combinations for this pillar. But, according to the rankings in Table 1, the sequence of animal symbols chosen is the best possible. The chance of this occurring randomly is 1 in 1728. To dispute this, one would need to find many other combinations of animal symbols that are a better match to the constellations than those that actually appear on Pillar 2.
3. The chance of choosing the animal symbol that represents the constellation at the peak intensity of the Taurids, the fox, is simply 1 in 12, as there are currently 12 animal symbols known.

Multiplying all these probabilities together gives a chance of 1 in 300 thousand that Göbekli Tepe does not implicate the Taurid meteor stream in the Younger Dryas impact event.

**Tables**

| Symbol | Asterism | Rank |
|---|---|---|
| 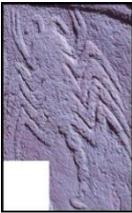 Scorpion | 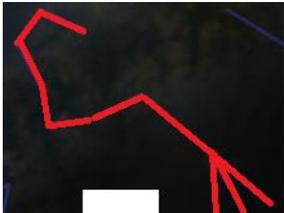 Scorpius | 1 |
| 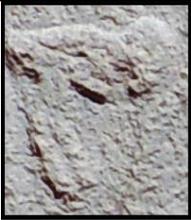 Bending bird | 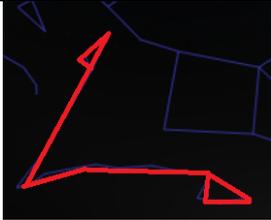 Pisces | 1 |
| 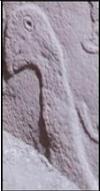 Duck/goose | 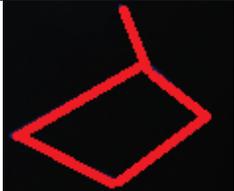 Libra | 1 |
| 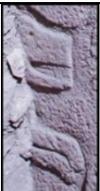 Dog/wolf? | 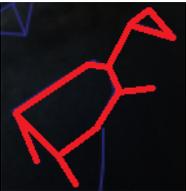 Lupus | 1 |
| 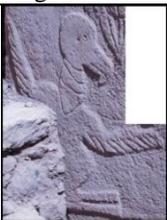 Eagle/vulture | 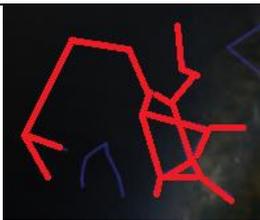 Sagittarius | 1 |
| 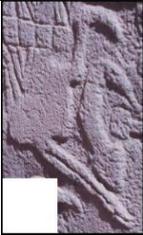 Bending bird with fish | 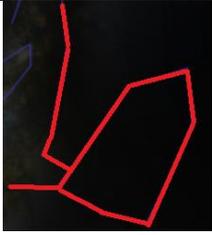 Ophiuchus | 1 |
| 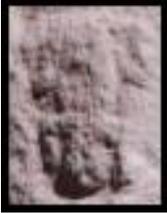 Bear | 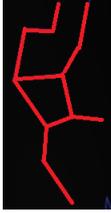 Virgo | 1 |

| Symbol | Asterism | |
|---|---|---|
| 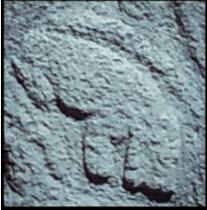 Charging ibex/gazelle | 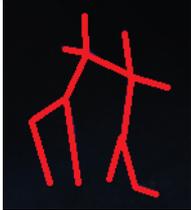 Gemini | =2 with lion/ leopard |
| 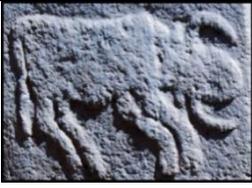 Aurochs | 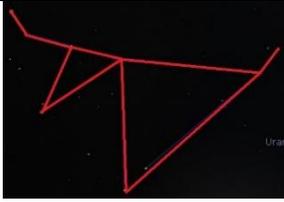 Capricornus | 1 |
| 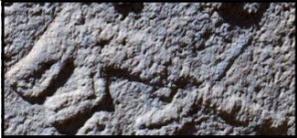 Fox | 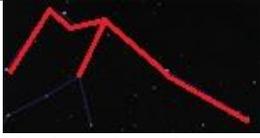 Northern Aquarius | 1 |

Table 1. Animal symbol – asterism associations identified at Göbekli Tepe.

| Symbol | Asterism |
|---|---|
| 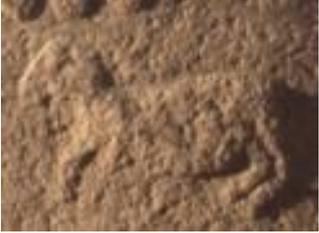 Ram | 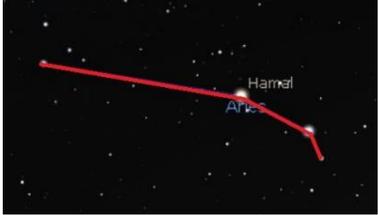 Aries |
| 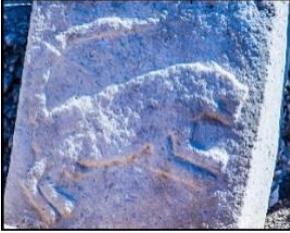 Pouncing lion/leopard | 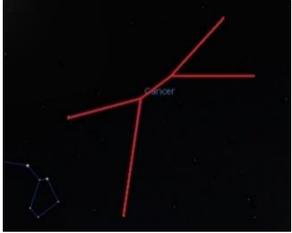 Cancer |

Table 2. Animal symbol – asterism associations deduced from Çatalhöyük.

| Symbol | Asterism |
|---|---|
| 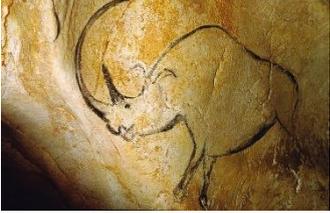 Rhinoceros at Chauvet (from 'Inocybe' via French Wikipedia) | 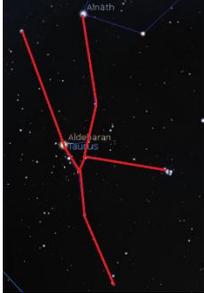 Taurus |
| 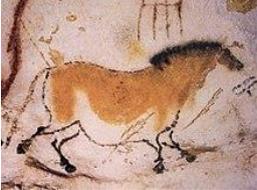 Horse at Lascaux (By 'Ownwork' via Wikipedia) | 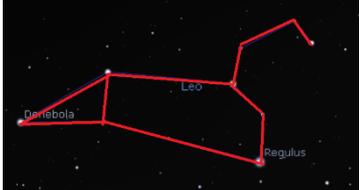 Leo |

Table 3. Animal symbol – asterism associations identified at Lascaux.

| Sample | C14 BP | Cal14 BC | Cal14 1σ (yr) | solstice/ equinox | centre BC | symbol | separation | notes |
|---|---|---|---|---|---|---|---|---|
| Altamira GifA 91178 | 13,570 +- 190 | 14410 | 280 | summer | 14350 | bison | 60 | |
| Altamira GifA 91179 | 13,940 +- 170 | 14940 | 270 | summer | 14350 | bison | 590 | |
| Altamira GifA 91181 | 14330 +-190 | 15480 | 260 | summer | 14350 | bison | 1130 | |
| Niaux GifA 91319 | 12890 +- 160 | 13470 | 240 | summer | 14350 | bison | 880 | |
| El castillo GifA 91172 | 12910 +- 180 | 13490 | 270 | summer | 14350 | bison | 860 | |
| Covaciella GifA 95281 | 14060 +- 140 | 15160 | 230 | summer | 14350 | bison | 810 | |
| Covaciella GifA 95364 | 14260 +- 130 | 15410 | 190 | summer | 14350 | bison | 1060 | |
| Chauvet GifA 95132 | 32410 +- 720 | 34510 | 930 | winter | 34200 | rhino | 310 | |
| Chauvet GifA 95133 | 30790 +- 600 | 32770 | 550 | winter | 34200 | rhino | 1430 | same as 95132 |
| average | | 33640 | 1070 | winter | 34200 | rhino | 560 | |
| Chauvet GifA 95126 | 30940 +- 610 | 32930 | 580 | winter | 34200 | rhino | 1270 | |
| Chauvet Gifa 95128 | 30340 +- 570 | 32400 | 470 | autumn | 32700 | bison | 300 | |
| Chauvet GifA 13034 | 31950 +- 460 | 33870 | 490 | spring | 34000 | horse | 130 | |
| Chauvet GifA 96065 | 30230 +- 530 | 32320 | 430 | autumn | 32700 | bison | 380 | |
| Chauvet GifA 11126 | 28170 +- 730 | 30170 | 770 | spring | 28750 | rhino | 1420 | |
| Chauvet GifA 13134 | 31830 +- 450 | 33760 | 510 | spring | 32350 | lion | 1410 | |
| Chauvet GifA 13105 | 31490 +- 430 | 33420 | 440 | autumn | 32700 | bison | 720 | |
| La Pileta GifA 98162 | 20310 +- 350 | 22540 | 460 | spring | 21200 | bison | 1340 | |
| Urdiales GifA 11454 | 12750 +- 110 | 13240 | 190 | summer | 14350 | bison | 1110 | |
| La Garma GifA 102581 | 13780 +- 150 | 14720 | 240 | summer | 14350 | bison | 370 | |
| Stadel figurine | 35185 +- 270 | 37800 | 340 | winter | 38650 | lion | 850 | |
| Hohle Fels figurine | 30500 +- 500 | 32500 | 420 | spring | 33950 | horse | 1450 | |
| Cosquer GifA 96069 | 26250 +- 350 | 28560 | 410 | winter | 27100 | bison | 1460 | |
| Cosquer GifA 95195 | 27350 +- 430 | 29310 | 370 | winter | 27100 | bison | 2210 | same as 96069 |
| Cosquer GifA 14157 | 26240 +- 270 | 28590 | 380 | winter | 27100 | bison | 1490 | same as 96069 |
| average | | 28820 | 470 | winter | 27100 | bison | 1720 | |
| Cosquer GifA 14159 | 18200 +- 110 | 20110 | 340 | spring | 21200 | bison | 1090 | |
| Cosquer GifA 14160 | 20120 +-510 | 22310 | 640 | spring | 21200 | bison | 1110 | |
| Cosquer GifA 92418 | 19200 +- 240 | 21200 | 300 | autumn | 20950 | lion | 250 | |
| Cosquer GifA 98186 | 19720 +-210 | 21800 | 410 | autumn | 23250 | horse | 1450 | |
| Cosquer GifA 14164 | 19890 +- 130 | 21980 | 330 | autumn | 23250 | horse | 1270 | |
| Catalhoyuk aurochs | | 7000 | - | autumn | 7400 | bison | 400 | |
| Catalhoyuk bear | | 7000 | - | summer | 5800 | bear | 1200 | |

Table 4. Radiocarbon data (15-22) compared with zodiacal measurements using Stellarium. Radiocarbon data is calibrated using the Calib704 software and the IntCal13 calibration curve. No distinction is made between bison and aurochs, or lion and leopard etc.

**Acknowledgments**
NA

**Ethical Statement**
NA

**Funding Statement**
NA

**Data Accessibility**
*NA*

**Competing Interests**
*We have no competing interests.*

**Authors' Contributions**
Alistair Coombs was instrumental in relating the symbols at Çatalhöyük to those at Göbekli Tepe. He also assisted with drafting the final manuscript. Martin Sweatman completed the rest of the work.